\documentclass[sigconf, nonacm]{acmart}
\AtBeginDocument{%
  }

\usepackage{subcaption} % 用于子图的标题  
\usepackage{multirow} % 用于表格
\usepackage[normalem]{ulem}
\usepackage{xcolor}
\definecolor{custompurple}{RGB}{155, 89, 182}
\definecolor{customblue}{RGB}{52, 152, 219}

\begin{document}

%%
%% The "title" command has an optional parameter,
%% allowing the author to define a "short title" to be used in page headers.
\title[Adapting AI to the Moment]{Adapting AI to the Moment: Understanding the Dynamics of Parent–AI Collaboration Modes in Real-Time Conversations with Children}
%%
%% The "author" command and its associated commands are used to define
%% the authors and their affiliations.
%% Of note is the shared affiliation of the first two authors, and the
%% "authornote" and "authornotemark" commands
%% used to denote shared contribution to the research.
\author{Yu Mei}
\authornote{These authors contributed equally to this work.}
\affiliation{%
  % \department{Department of Computer Science and Technology}
  \institution{Tsinghua University}
  \city{Beijing}
  \country{China}
}
\email{meiy24@mails.tsinghua.edu.cn}

\author{Ziyao Zhang}
\authornotemark[1]
\affiliation{%
  % \department{Department of Computer Science and Technology}
  \institution{Tsinghua University}
  \city{Beijing}
  \country{China}
}
\email{ziyao-zh24@mails.tsinghua.edu.cn}

\author{Qingyang Wan}
\affiliation{%
  % \department{Academy of Arts \& Design}
  \institution{Tsinghua University}
  \city{Beijing}
  \country{China}
}
\email{wanqy23@mails.tsinghua.edu.cn}

\author{Shiyi Wang}
\affiliation{%
  % \department{Academy of Arts \& Design}
  \institution{Tsinghua University}
  \city{Beijing}
  \country{China}
}
\email{shiyi-wa23@mails.tsinghua.edu.cn}

\author{Ge Wang}
\affiliation{%
  % \department{Siebel School of Computing and Data Science}
  \institution{University of Illinois at Urbana-Champaign}
  \city{Champaign, Illinois}
  \country{United States}
}
\email{wangge@illinois.edu}

\author{Jie Cai}
\affiliation{%
  % \department{Department of Computer Science and Technology}
  \institution{Tsinghua University}
  \city{Beijing}
  \country{China}
}
\email{jie-cai@mail.tsinghua.edu.cn}

\author{Chun Yu}
\affiliation{%
  % \department{Department of Computer Science and Technology}
  \institution{Tsinghua University}
  \city{Beijing}
  \country{China}
}
\email{chunyu@tsinghua.edu.cn}

\author{Yuanchun Shi}
\affiliation{%
  % \department{Department of Computer Science and Technology}
  \institution{Tsinghua University}
  \city{Beijing}
  \country{China}
}
\email{shiyc@tsinghua.edu.cn}

%%
%% By default, the full list of authors will be used in the page
%% headers. Often, this list is too long, and will overlap
%% other information printed in the page headers. This command allows
%% the author to define a more concise list
%% of authors' names for this purpose.
\renewcommand{\shortauthors}{Mei et al.}

%%
%% The abstract is a short summary of the work to be presented in the
%% article.
\begin{abstract}
% Parent-child conversations play a crucial role in children's development, yet are challenging for parents. Prior work on AI support has primarily focused on macro-level tasks, but has underestimated the moment-to-moment adaptations that such dynamic conversations demand. To explore these micro-level dynamics, we conducted a co-design study with eight parents and developed COMPASS, which enables parents to combine scaffolding functions in real time flexibly during parent-child conversations. Using COMPASS as a research  probe, we conducted a lab-based user study with 21 parent-child pairs. We found that parents developed preferred modes of AI collaboration, which evolved over time and shifted systematically with contextual factors. We further found that parents employed parent-oriented, child-oriented, and relationship-oriented to engage with AI support. This work advances understanding of dynamic human-AI collaboration in parenting contexts, providing design implications for future flexible, context-adaptive parental support systems.

% Parent-AI collaboration to support real-time conversation with a child is challenging due to the sensitivity and open-ended nature of children.  
% Existing systems often simplified the collaboration mode and provided limited support for adapting AI to continuously changing conversational contexts.

Parent-AI collaboration to support real-time conversations with children is challenging due to the sensitivity and open-ended nature of such interactions. Existing systems often simplify collaboration into static modes, providing limited support for adapting AI to continuously evolving conversational contexts. To address this gap, we systematically investigate the dynamics of parent-AI collaboration modes in real-time conversations with children. We conducted a co-design study with eight parents and developed COMPASS, a research probe that enables flexible combinations of parental support functions during conversations. Using COMPASS, we conducted a lab-based study with 21 parent-child pairs. We show that parent–AI collaboration unfolds through evolving modes that adapt systematically to contextual factors. We further identify three types of parental strategies---parent-oriented, child-oriented, and relationship-oriented---that shape how parents engage with AI. These findings advance the understanding of dynamic human-AI collaboration in relational, high-stakes settings and inform the design of flexible, context-adaptive parental support systems.
\end{abstract}

%%
%% The code below is generated by the tool at http://dl.acm.org/ccs.cfm.
%% Please copy and paste the code instead of the example below.
%%


%%
%% Keywords. The author(s) should pick words that accurately describe
%% the work being presented. Separate the keywords with commas.
%% A "teaser" image appears between the author and affiliation
%% information and the body of the document, and typically spans the
%% page.
\begin{teaserfigure}
\centering
\includegraphics[width=1.0\columnwidth]{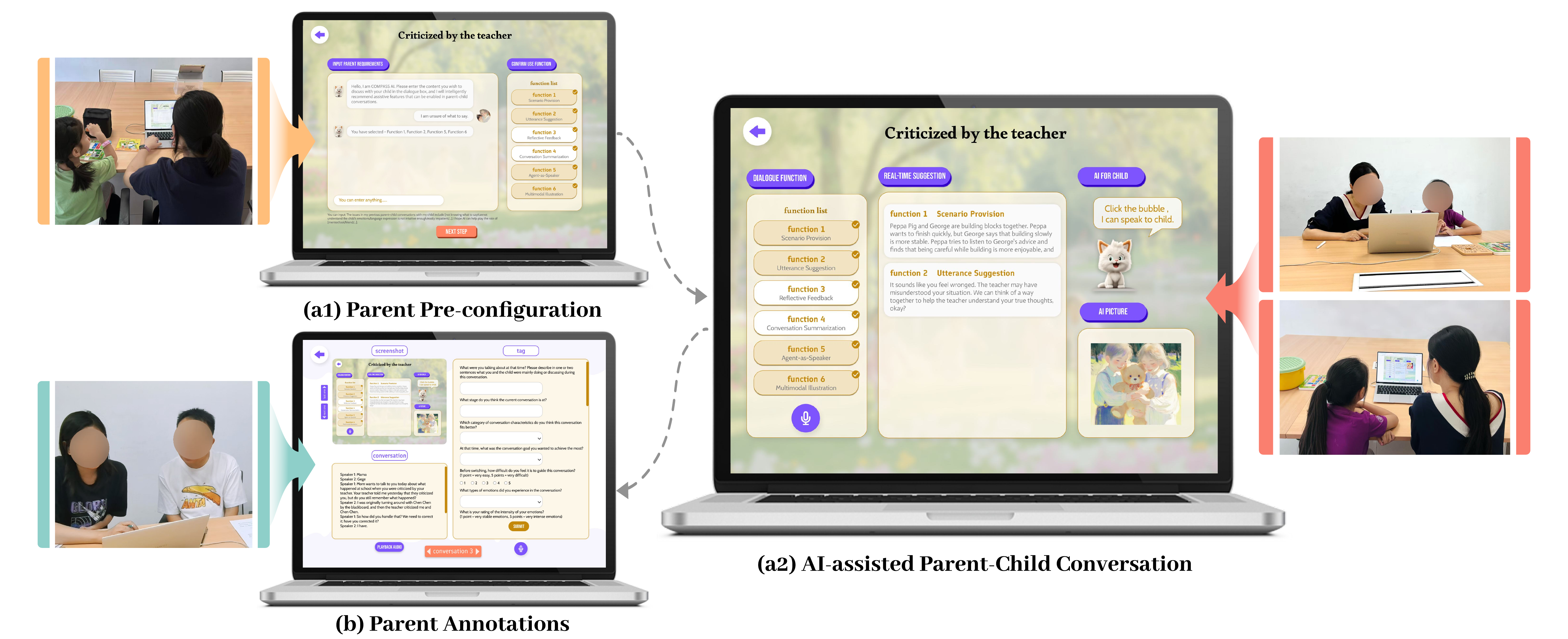}
\caption{Main screens of the COMPASS research probe: (a1) Parent pre-configuration interface for selecting collaborative functions; (a2) AI-assisted parent-child conversation interface; (b) Parent annotation interface for post-conversation annotations.}
\label{figure:User Interface}
\end{teaserfigure}

% \received{20 February 2007}
% \received[revised]{12 March 2009}
% \received[accepted]{5 June 2009}

%%
%% This command processes the author and affiliation and title
%% information and builds the first part of the formatted document.
\maketitle

\section{Introduction}
% Parent-AI collaboration in \textit{real-time conversational settings with children} presents unique challenges. Unlike task-oriented interactions with predefined goals and structures, these settings involve sensitive and open-ended exchanges in which parents must integrate AI support in response to evolving conversational contexts \cite{driscoll2026understanding, morelen2012real, mazzone2017parents}. As a result, parents are engaged in continuous decision-making, determining how and when AI should participate throughout the interaction.
% These challenges are further amplified by the dynamic and non-linear nature of conversational interactions, particularly those involving children. Topics, emotional tone, and engagement levels can shift rapidly within a single interaction \cite{lin2023assessment, overbeek2022quality, leclere2014synchrony}. As these contextual factors evolve, parents' needs for AI support also change accordingly. For example, parents may rely on AI for idea generation when fatigued \cite{dietz2024contextq}, but expect emotional comfort or mediation in more sensitive situations \cite{theofanopoulou2022exploring}. Such rapid and context-dependent shifts make it difficult to predefine effective collaboration strategies, requiring AI support to adapt continuously during interaction.

Parent-AI collaboration in \textit{real-time conversational settings with children} presents unique challenges. Unlike task-oriented interactions with predefined goals and structures, these settings involve sensitive and open-ended exchanges in which conversational contexts can shift rapidly within a single interaction \cite{driscoll2026understanding, morelen2012real, lin2023assessment, overbeek2022quality}. As a result, parents' needs for AI support vary dynamically, ranging from seeking idea generation when fatigued \cite{dietz2024contextq} to expecting emotional mediation in sensitive moments \cite{theofanopoulou2022exploring}. This variability requires parents to continuously decide when and how AI should participate, making it difficult to predefine stable collaboration strategies.

However, existing AI systems for conversational support typically rely on a limited set of static collaboration modes, such as generating prompts or acting as a conversational agent \cite{ho2025set, zhang2022storybuddy, liu2024compeer, srinivasan2021snowy}. While these approaches demonstrate the potential of AI to assist in conversations, they assume that the set of collaboration modes can be predefined for the entire interaction, providing limited support for adapting AI to evolving conversational contexts.

Despite the growing interest in AI-supported parenting tools, little is known about how parents' preferred collaboration modes with AI emerge and evolve during ongoing conversations. To address this gap, we investigate how parents configure AI involvement in ongoing interactions. We introduce the concept of \textit{collaboration modes} to describe how parents configure and engage AI functions during interactions.

We therefore explore the following research questions:
\begin{itemize}
    \item \textbf{RQ1}: What modes of AI collaboration do parents prefer in parent-child conversations?
    \item \textbf{RQ2}: How do contextual factors influence parents' preferred modes of AI collaboration in parent-child conversations?
    \item \textbf{RQ3}: What strategies do parents use to engage with AI support during parent-child conversations?
\end{itemize}

To answer these questions, we conducted a co-design study with eight parents to elicit parental needs for AI support in parent-child conversations. We revealed six functional needs (F1-F6) and two key requirements for collaborating with AI: flexible decision authority and minimal mode-switching costs. We then developed COMPASS, a \underline{co}llaborative \underline{m}ediation and \underline{pa}rental \underline{s}caffolding \underline{s}ystem, as a research probe to explore parent-AI collaboration in conversation. COMPASS allows parents to flexibly combine parental support functions during conversations, enabling us to examine how their strategies adapt to shifting collaboration needs.

Using COMPASS as a research probe, we conducted a lab-based study with 21 parent-child pairs in semi-structured conversational tasks. We found that parents developed preferred collaboration modes with AI, which moved towards a more partner-like collaboration mode and converged into recurring patterns (RQ1). Parents adjusted these modes systematically based on emotional intensity, emotion type, conversation stage, parental goals, and conversation scale and structure (RQ2). Additionally, parents adopted distinct strategies to engage with AI support across three dimensions: parent-oriented, child-oriented, and relationship-oriented (RQ3).

Our study contributes to HCI in four ways. First, we provide empirical insights into the dynamic nature of parent-AI collaboration, showing how collaboration modes evolve over time, vary with contextual factors, and are enacted through parental strategies. Second, we extend Parental Goal Theory by grounding its three established categories in our empirical findings of AI-assisted parenting. Third, we advance the understanding of human-AI collaboration in relational, high-stakes domains. Finally, we offer design implications for parental support systems that enable flexible, context-adaptive human-AI collaboration.
\section{Related Work}

\subsection{Parent–Child Conversations in HCI}
Parent–child communication is critical to early childhood development, supporting children's cognitive, social, and emotional growth through everyday interactions \cite{vygotsky2012thought, hart1997meaningful, grotevant2001developing, mclean2007selves, carson1999family, fitzpatrick2005family}. As children's first social environment, families position parents as primary agents of socialization. In these interactions, parents continuously guide, respond to, and adapt to children's needs, shaping the quality and outcomes of communication.

Prior HCI research has explored supporting parent–child conversations from multiple perspectives \cite{zapf2023systematic}, including (1) content provisioning (e.g., prompts or games) \cite{chan2017wakey}; (2) process guidance that analyzes conversation dynamics and provides feedback (e.g., tone or emotional cues) \cite{choi2025aacesstalk, dietz2024contextq}; and (3) third-party mediators (e.g., social robots) that restructure interaction \cite{ho2024s, ho2025set, seo2025enhancing}. While these approaches provide valuable support, they largely assume stable interaction, limiting their ability to adapt to dynamic conversations.
% \jc{above, we should say hci research focus on prenting support or scaffold from various perspecives such as xxxxx for support parents, xxx for scaffolding parents,  but these mainly consider it is statics, not very context senitive?}

In practice, parent-child conversations are highly dynamic and context-sensitive. Parents often balance multiple, shifting goals---such as guidance, protection, reciprocity, and control---while responding to children's changing cognitive, emotional, and attentional states \cite{krevans1996parents, knafo2003parenting, grusec1994impact, morelen2012real, denham1998emotional, flavell1994cognitive, posner2007educating}. This requires parents to continuously adjust how they engage in conversations.

Prior work has explored parent–AI collaboration in conversational settings \cite{ho2023designing, zhang2022storybuddy}. However, how such collaboration evolves in response to changing goals and contexts has not been systematically examined. Our work addresses this gap by investigating the dynamics of parent–AI collaboration in real-time conversations and showing how these dynamics can inform the design of adaptive AI support for parent–child communication.

\subsection{AI Systems for Parent-Child Interaction}
Prior AI systems for parent-child education typically predefine parent-AI roles before interaction. Existing work identifies three collaboration modes. In the \textit{parent-led, AI-assisted} mode, AI supports parents' emotional guidance, story co-creation, and content selection through analysis, generation, and feedback \cite{shen2025easel, xu2025accompany, kim2025bridging, zhao2025youthcare}. In the \textit{AI-led, parent-assisted} mode, AI acts as the primary interactive agent while parents supervise or intervene when needed \cite{lau2025exploring, zhang2025exploring}. A third mode emphasizes \textit{joint collaboration}, positioning AI as a conversational partner in multi-turn interactions \cite{viswanathan2025interaction, shang2025learner}.

While these systems explore different role allocations, they primarily focus on static collaboration structures \cite{li2025learning, wang2026division}, with less attention to how parents can adapt AI involvement during ongoing interactions. In practice, parent-child conversations are highly dynamic, with emotions, goals, and engagement shifting in real time \cite{morelen2012real, bradley2015ebb}. Supporting parents in responding to these changes requires more in situ control over AI behavior.

Recent work has begun to explore more flexible forms of parental participation. For example, parents express a desire to adjust AI content generation, interaction depth, and collaboration structures according to contextual factors such as children's learning stages, time availability, and energy levels \cite{ho2025set, zhang2022storybuddy, ho2023designing, ho2024s}. Other studies introduce configurable parental control tools that support differentiated permissions, interaction modes, and intervention intensity across developmental stages \cite{dumaru2025one, yang2023towards}.

However, these approaches largely assume relatively static collaboration modes throughout the interaction, offering limited support for adapting AI behavior to rapidly changing conversations. In addition, prior work provides limited theoretical grounding for understanding how and why parents adjust collaboration with AI in real time. Parental Goal Theory \cite{hastings1998parenting} suggests that parents simultaneously pursue parent-, child-, and relationship-oriented goals that shift during interaction. Building on this perspective, we extend the theory to explain how parents regulate collaboration with AI under changing contextual factors.

\subsection{Collaboration Dynamics in Human-AI Interaction}
HCI research has long examined how humans and AI share roles in collaboration. Prior work often frames AI through static roles. As a tool or assistant, AI improves efficiency while humans retain primary control \cite{wu2022ai, wang2019human, lee2022coauthor}. As a coach or mentor, AI provides guidance without taking over decisions \cite{arakawa2024coaching, fadhil2019coachai}. As a teammate or partner, AI collaborates with users toward shared goals through complementary capabilities \cite{zhang2021ideal, brandt2009two, salikutluk2024evaluation}. However, these roles are typically fixed at design time, limiting collaboration to stage-based arrangements and overlooking interaction dynamics.

Recent work frames collaboration as ``agency negotiation'', where initiative and control shift with task stages and contextual factors \cite{masson2024directgpt}. Adjustable autonomy can improve perceived usefulness, competence, and satisfaction \cite{houde2025controlling, zhang2023follower}, and enhance team performance when AI adapts during collaboration \cite{salikutluk2024evaluation}. While prior work highlights interaction-level dynamics, most systems examine these dynamics at coarse-grained task units (e.g., storytelling or search sessions) \cite{zhang2022storybuddy, sharma2024generative}, with limited work systematically capturing interaction dynamics or validating them through system design.

Recent systems explore finer-grained collaboration mechanisms, including proactive intervention timing \cite{oh2024better, kuang2024enhancing, zhu2025autopbl}, proactive task collaboration \cite{pu2025assistance, mei2025interquest}, shared representations for alignment \cite{jeong2025conversation, zhang2025ladica}, and role fluidity at the micro-task level \cite{zhou2024understanding, liu2025proactive, wu2025negotiating}.

However, these studies primarily address general task collaboration and rarely examine parent-child communication, a setting characterized by multiple concurrent goals \cite{krevans1996parents, knafo2003parenting}, strong emotional fluctuations \cite{denham1998emotional}, and asymmetric roles \cite{elbers2004conversational}. Moreover, they lack a theoretical account explaining how users select collaboration modes under such conditions.

Building on this literature, we examine collaboration dynamics in emotionally sensitive parent–child conversations, tracing how parents adjust control and support in real time and extending prior frameworks from macro-level autonomy to micro-level interaction.
\section{Formative Study}
To understand parents' needs and preferred modes of collaborating with AI in parent–child conversations, we conducted a formative co-design study to inform the design of our research probe.

\subsection{Setup}
\subsubsection{Protocol}
The study consisted of a pre-task and a co-design session. 
Before the session, parents engaged in a 10-minute conversation with their child about daily challenges (e.g., peer conflict, teacher criticism), audio-recorded the interaction, and shared it with the research team.

\begin{figure}[htb] 
\centering
\includegraphics[width=\columnwidth]{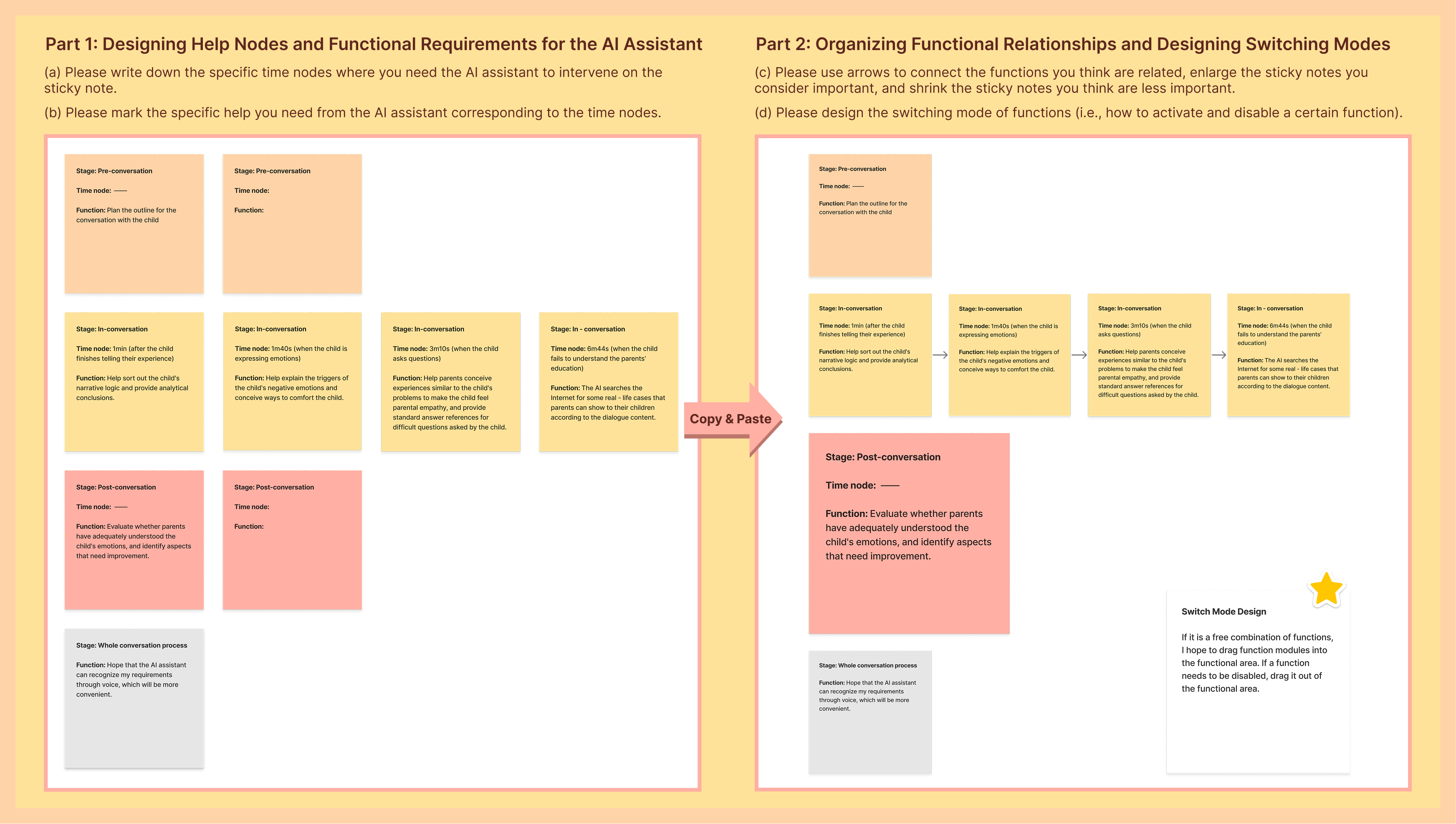}
\caption{Illustration of the co-design session.}
\label{figure:Codesign Procedure}
\end{figure}

During the co-design session, parents reviewed their recordings with two HCI researchers and identified moments where AI support could be helpful using an online whiteboard (\autoref{figure:Codesign Procedure}). Conversations were structured into four phases (before, during, after, and throughout), and parents annotated desired AI support for each phase. They then organized, prioritized, and combined these functions, and discussed feature-switching preferences and usage requirements. Participants were encouraged to think aloud, and facilitators used follow-up questions to elicit detailed reflections.

\subsubsection{Participants}
\label{Section: Formative Study Participants}
We recruited eight parents (P1-P8) of children aged 6-8 via social media. This age group was selected because children in the early primary school years still rely on parental guidance \cite{sroufe2005development, gottman2013meta} while developing conversational abilities that allow them to engage in meaningful conversations \cite{berman2013relating, mccarthy1943language}.

Participants included eight parents (5 fathers, 3 mothers; $M = 38.62$, $SD = 4.18$) and their children (2 boys, 6 girls; $M = 6.50$, $SD = 0.71$). Sessions were conducted via video conferencing ($\sim$1 hour) and facilitated by two HCI researchers. All participants provided informed consent and received \$10 or equivalent compensation.

\subsection{Findings}
\subsubsection{AI Support Functions Aligned with Parental Goals}
\label{section:AI Support Functions Aligned with Parental Goals}
From the co-design sessions, we identified six frequently requested AI support functions (F1-F6), which we group into two categories based on Parental Goal Theory \cite{dix1991affective}. 

\textbf{Parent-focused goals} include Scenario Provision (F1), Utterance Suggestion (F2), and Reflective Feedback (F3), supporting parents in framing topics, expressing themselves, and reflecting on their strategies. Parents preferred contextualized storytelling over fixed scripts (F1; P6:``\textit{making it easier for children to understand and accept}''), concise and directly usable utterances at critical moments (F2; P8:``\textit{could read aloud directly}''), and non-judgmental, heuristic feedback (F3; P2:``\textit{I need it to inspire me on how to think}'').

\textbf{Relationship-focused goals} include Conversation Summarization (F4), Agent-as-Speaker (F5), and Multimodal Enrichment (F6), supporting mutual understanding and engagement. Parents expected real-time summaries to track emotions and perspectives (F4; P1:``\textit{so that everyone can understand each other}''), occasional AI participation as a lightweight third party when conversations stalled (F5; P4:``\textit{third-party friend}''), and multimodal content (e.g., images) to help children understand the current conversation and sustain engagement (F6; P5:``\textit{he will be more willing to continue the conversation}'').

Child-focused goals were excluded, as our study focused on parent-centered collaboration needs in conversations. Details of the six functions (F1–F6), including their purposes and use contexts, are provided in \autoref{appendix:Details of the Six Parent Support Functions}.

\subsubsection{Flexible Function Combination with Retained Decision Authority}
In the co-design sessions, parents emphasized the need to dynamically adapt support functions to rapidly changing conversational contexts.

Preferences varied. Some parents preferred full manual control, advocating ``\textit{free combination}'' to maintain flexibility and natural interaction: ``\textit{The freer, the better... the more fixed its feedback format is, the less natural it feels.}'' (P4). Some parents favored preconfigured function combinations to reduce effort: ``\textit{I think fixed system functions would be better, so I don't have to switch frequently.}'' (P7)

Most parents preferred a hybrid approach, where AI suggests an initial set of combinations while leaving final control to parents. P6 noted, ``\textit{You can offer suggestions for them to adopt},'' and P8 suggested ``\textit{it could prompt whether I need certain functions and I can just select `yes'.}'' Across these views, parents consistently stressed retaining decision authority in choosing functions, as needs vary across situations and conversations are inherently unpredictable.

\subsubsection{Minimizing Interactional Costs of Collaboration Mode Switching}
Parents noted that frequent manual switching could disrupt conversational flow, suggesting a preference for minimally intrusive interaction (P7: ``\textit{when talking to children...keeping a fixed mode might be better}'').

To reduce switching costs, parents proposed more seamless interaction modalities. Voice control was seen as effective when typing is inconvenient (P5: ``\textit{Voice commands... more convenient and frees my hands}''), while many favored lightweight, intuitive interfaces such as one-click toggles (P2: ``\textit{Clicking is best---it matches our usual habits.}''). Some also suggested drag-and-drop interactions for flexible configuration (P6).

Overall, parents preferred switching mechanisms that are low-interference, predictable, and easily reversible, enabling real-time adaptation without disrupting conversational continuity.
\section{COMPASS Design and Implementation}
Building on insights from the formative study, we designed and implemented COMPASS. Rather than functioning as a full intervention system, COMPASS is intended to investigate how parents' collaboration needs and behaviors with AI evolve during such conversations. To preserve natural decision-making, we minimized system-induced bias that could influence parents' collaboration choices. To support analysis of parent-AI collaboration dynamics, we incorporated an annotation interface to capture in-situ reflections and strategies.

\begin{figure*}[] 
\centering
\includegraphics[width=0.8\textwidth]{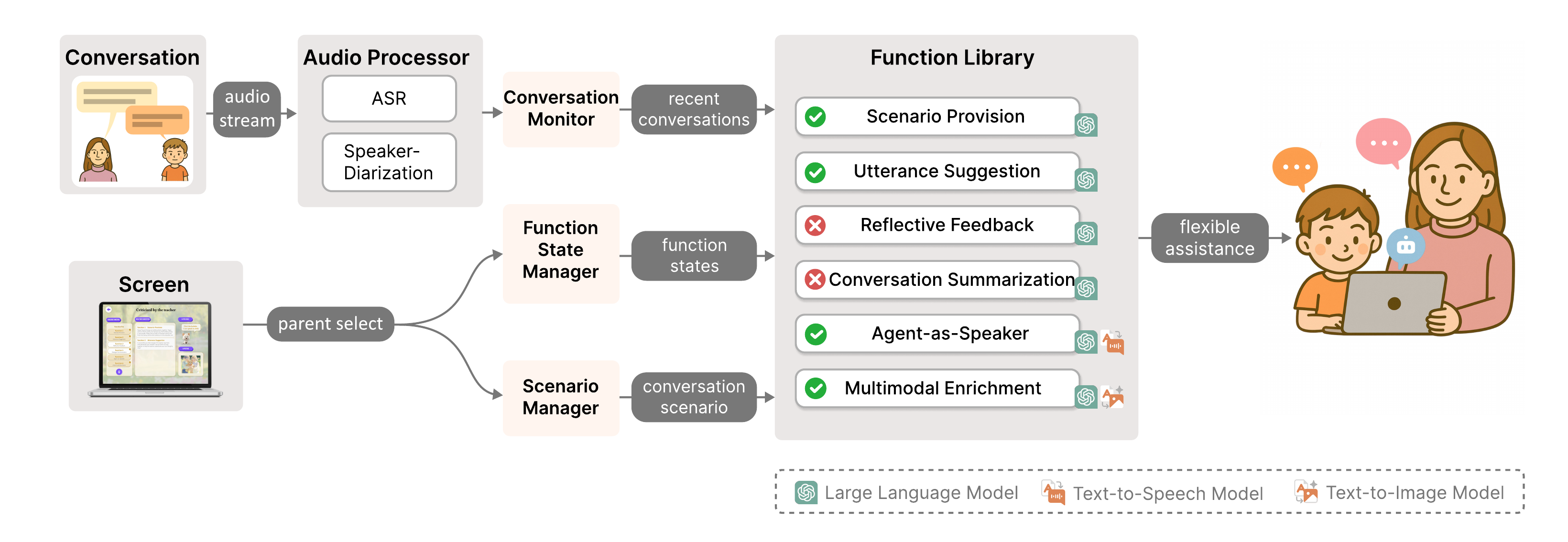}
\caption{Overview of the COMPASS research probe workflow.}
\label{figure:System Architecture}
\end{figure*}

\subsection{User Interface Design}
The COMPASS interface consists of two components: \textit{Parent–Child Conversation} and \textit{Parent Annotation} (\autoref{figure:User Interface}).

\subsubsection{Parent-Child Conversation Interface}
\label{Parent-Child Conversation Interface}
This interface helps parents conduct topic-based conversations with their children. Based on the findings from the formative study, we derived a set of design requirements (\autoref{tab:finding-design-mapping}).

\begin{table}[]
\caption{Mapping from formative study findings to COMPASS design features.}
\centering
\resizebox{\columnwidth}{!}{
\begin{tabular}{ll}
\toprule[1pt]
\multicolumn{1}{c}{\textbf{Formative Study Finding}}                                                           & \multicolumn{1}{c}{\textbf{COMPASS Design Feature}}                                                                                           \\ \hline
\begin{tabular}[c]{@{}l@{}}AI support functions aligned \\ with parental goals\end{tabular}                   & \begin{tabular}[c]{@{}l@{}}Six modular functions in the \\ Parent-Child Conversation Interface\end{tabular}                                       \\ \hline
\begin{tabular}[c]{@{}l@{}}Flexible function combination \\with retained decision authority\end{tabular} & \begin{tabular}[c]{@{}l@{}}Initial function set based on stated purpose; \\ parents can enable, disable, or reorder at any time.\end{tabular} \\ \hline
\begin{tabular}[c]{@{}l@{}}Minimizing interactional costs of \\ collaboration mode switching\end{tabular}      & \begin{tabular}[c]{@{}l@{}}Auto-update the active modules every 30 s; \\ enable/disable functions with a single click\end{tabular}            \\ \bottomrule[1pt]
\end{tabular}
}
\label{tab:finding-design-mapping}
\end{table}

To address the identified functional needs (F1–F6), COMPASS includes six modules: \textit{Scenario Provision}, \textit{Utterance Suggestion}, \textit{Reflective Feedback}, \textit{Conversation Summarization}, \textit{Agent-as-Speaker}, and \textit{Multimodal Enrichment}. Each module provides contextual support during the conversation.

Before the conversation, parents configure an initial set of functions (a1 in \autoref{figure:User Interface}). They can describe their communication needs in natural language, based on which COMPASS suggests a set of functions via LLM-based reasoning. Parents can also manually adjust the collaboration mode.

During the parent–child conversation (a2 in \autoref{figure:User Interface}), COMPASS provides support through the enabled modules. Parents can enable or disable modules at any time, enabling flexible adaptation to evolving interactions. To balance responsiveness and minimal disruption, continuous-support modules (F1-F4) are automatically updated every 30 seconds when enabled, while the remaining modules (F5-F6) are triggered on demand. Functional modules are presented in a compact list, enabling quick switching with minimal cognitive and operational effort.

\subsubsection{Parent Annotation Interface}
To examine the rationale behind parents' collaboration choices, COMPASS includes a post-conversation annotation interface (b in \autoref{figure:User Interface}). The conversation is segmented into slices based on moments when parents change the activation of one or more modules; consecutive changes are merged into a single slice if no conversation occurs between them.

\begin{figure}[htb] 
\centering
\includegraphics[width=1.0\columnwidth]{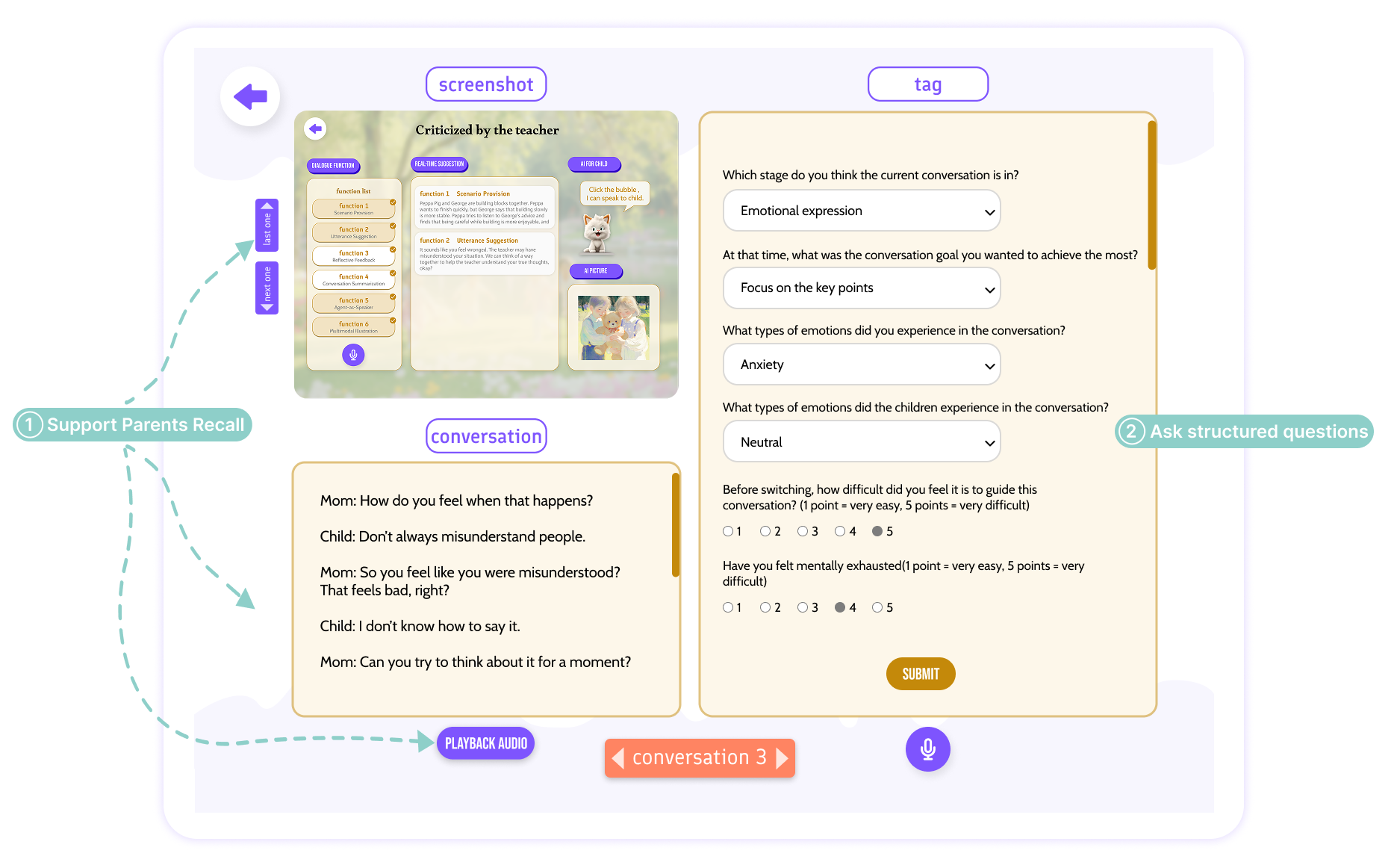}
\caption{Example parent annotation for a conversation slice.}
\label{figure:Annotation Example}
\end{figure}

For each conversation slice, COMPASS presents the transcript, snapshot, and audio replay to support recall. Parents are then asked to respond to structured questions about contextual factors during that slice, including their own state, the child's state, and interaction-related factors (see all rows labeled ``Annotator = Parent'' in Table 2). An example of a single annotation is shown in \autoref{figure:Annotation Example}.

\subsection{COMPASS Workflow}
We describe the workflow of COMPASS as a research probe supporting parent–child conversations (\autoref{figure:System Architecture}).

During the parent-child conversation, the \textit{Audio Processor} performs automatic speech recognition (ASR) and speaker diarization on the live audio stream. The resulting transcripts, annotated with speaker labels, are passed to the \textit{Conversation Monitor}, which maintains a sliding window of recent utterances.

Meanwhile, parents can enable or disable functions (F1-F6) through the \textit{Parent–Child Conversation} interface, which updates the \textit{Function State Manager} in real time. Parents may select a predefined topic from the \textit{Scenario Manager} to initiate a conversation. The scenario set includes eight situations identified in the formative study as challenging for children aged 6-8 (e.g., peer conflict). 

The current context, including recent utterances, function states, and the selected scenario, is passed to the \textit{Function Library} to generate context-aware support for the parent. The \textit{Function Library} contains six modules (Section~\ref{section:AI Support Functions Aligned with Parental Goals}) that operate independently. These modules leverage LLMs, text-to-speech, and text-to-image models to provide multi-channel support. 

Implementation details are provided in the \autoref{appendix:Implementation of the COMPASS Research Probe}.

\section{User Study}
We conducted a lab-based study with 21 parent-child pairs using COMPASS as a research probe to examine (1) parents' preferred AI-collaboration modes, (2) how conversational contextual factors shape these preferences, and (3) strategies parents use to engage with AI during parent–child conversations. The study was IRB-approved, and informed consent/assent was obtained from all participants.

\subsection{Participants}
We recruited 21 parent-child pairs via social media. Each pair received \$20 or equivalent compensation. Parents (P1-21) included 4 fathers and 17 mothers ($M=39.05$, $SD=4.02$). Children (C1-21) included 11 boys and 10 girls aged 6-8 ($M=7.00$, $SD=0.76$); the rationale for this age group is provided in Section~\ref{Section: Formative Study Participants}. Detailed demographics are provided in \autoref{appendix:Additional User Evaluation Details}.

All sessions were recorded (video, audio, screen) with anonymization. Throughout the study, system outputs were monitored remotely to prevent harmful content.

\subsection{Procedure}
Each parent-child pair participated in a continuous two-hour lab session using a touchscreen laptop.

\textbf{Introduction (25 min).} Parents were introduced to COMPASS and its six modular functions, followed by a 10-minute free conversation to familiarize themselves with the system. They were encouraged to explore different feature combinations and collaboration strategies.

\textbf{Parent–child conversation (20 min).} Parents and children discussed one of two randomly assigned challenging topics (e.g., peer conflict, teacher criticism), selected based on formative interviews with five parents. Parents interacted with the system while guiding their children.

\textbf{Co-annotation (45 min).} Using the Parent Annotation Interface, parents and an HCI researcher jointly reviewed conversation segments. Parents recalled interactions via replay, annotated contextual factors (\autoref{tab:contextual factors list}), and explained their decisions through researcher probing. Children stayed in a supervised adjacent area during this and the next session.
% Objective measures (e.g., turn-taking) were automatically computed or later coded, while subjective measures (e.g., cognitive load, emotion, engagement) were parent-reported.
% In designing the contextual factors for annotation, we drew on Parental Goal Theory \cite{dix1986social, hastings1998parenting} and classified them into three categories: parent-oriented, child-oriented, and relationship-oriented. Qeustions and optiona lists are from prior theories, including Conversation Analysis\cite{sacks1974simplest}, Speech Act Theory\cite{austin1975things}, Scaffolding Theory\cite{wood1976role, puntambekar2005tools}, Cognitive Load Theory\cite{sweller1998cognitive} and Discrete Emotion Theory\cite{izard2009emotion}. Please see appendix for details of the annotation metrics. Objective measures, such as conversation acts or turn-taking patterns, were computed automatically by PACEE or later coded by HCI researchers. Meanwhile, more subjective measures (e.g., cognitive load, emotion type, engagement) were annotated by parents.

\textbf{Final interview (15 min).} We conducted semi-structured interviews to elicit parents' overall strategies for engaging with AI.

\subsection{Data Analysis}
For quantitative analysis, we pre-processed COMPASS logs and questionnaire data. Numeric features (e.g., turn count, utterance length) were z-standardized. For each conversation slice, collaboration modes were encoded as six-digit binary vectors indicating the activation of each function. Questionnaire responses (e.g., stage, emotion type) were one-hot encoded. We then analyzed function usage over time and its correlations with contextual factors.

For qualitative analysis, we transcribed co-annotation and interview sessions and conducted thematic analysis \cite{braun2006using,terry2017thematic, braun2019reflecting}. Four HCI researchers performed iterative coding: initial open coding, collaborative codebook development with one participant (P1), and independent coding with consensus discussions. Codes were then clustered to identify recurring parental strategies and collaboration patterns. All analyses were conducted in Mandarin to preserve original meanings.
\section{Findings}

\subsection{Parents' Preferred Modes of AI Collaboration (RQ1)}

\begin{figure}[h]
    \centering
    \begin{minipage}[t]{0.20\textwidth}
        \centering
        \includegraphics[width=\textwidth]{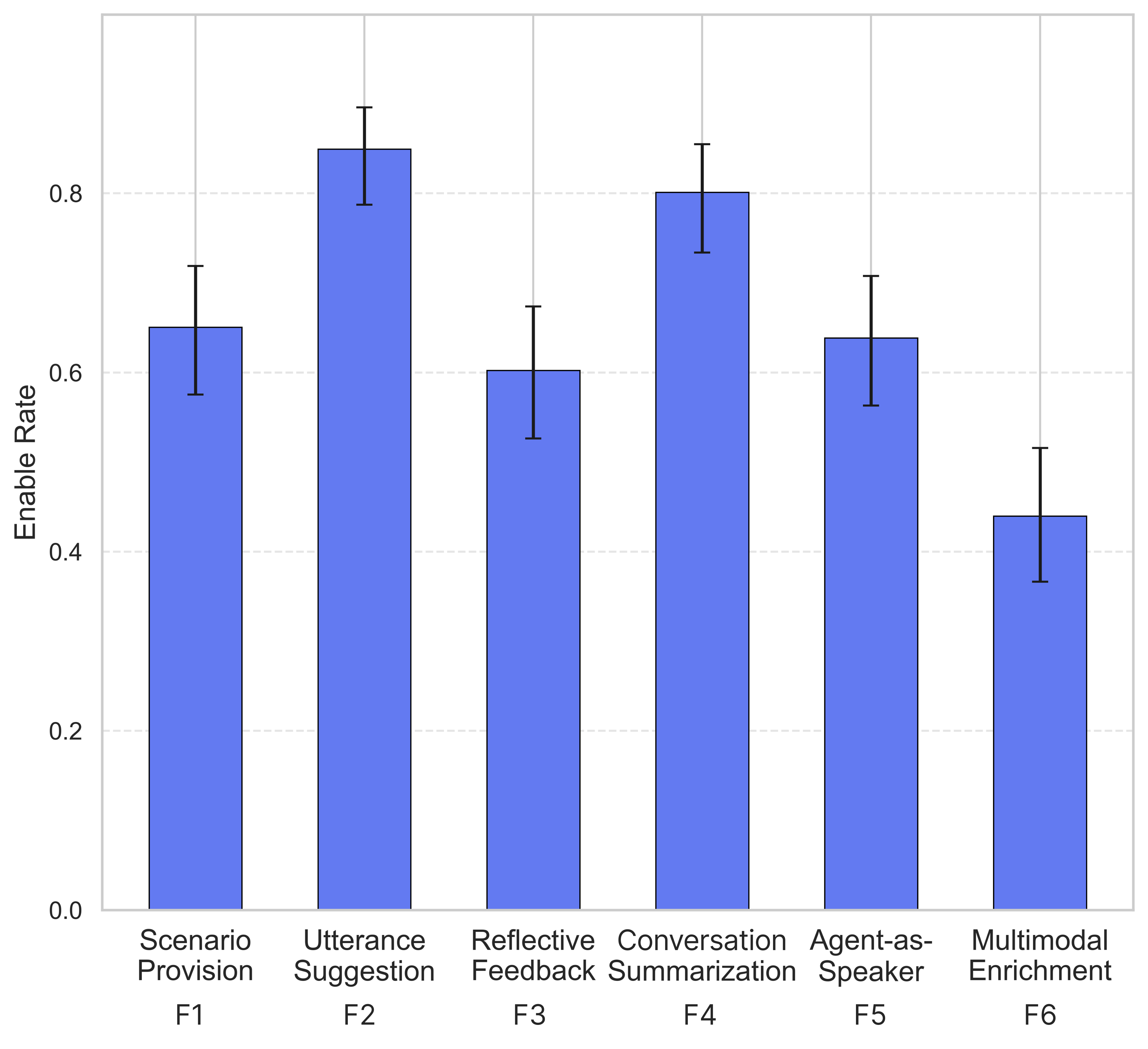} 
        \caption{Overall function enable rates (95\% CI).}  
        \label{figure:Function Enable Rates Bar Chart}
    \end{minipage}
    \hfill
    \begin{minipage}[t]{0.268\textwidth}
        \centering
        \includegraphics[width=\textwidth]{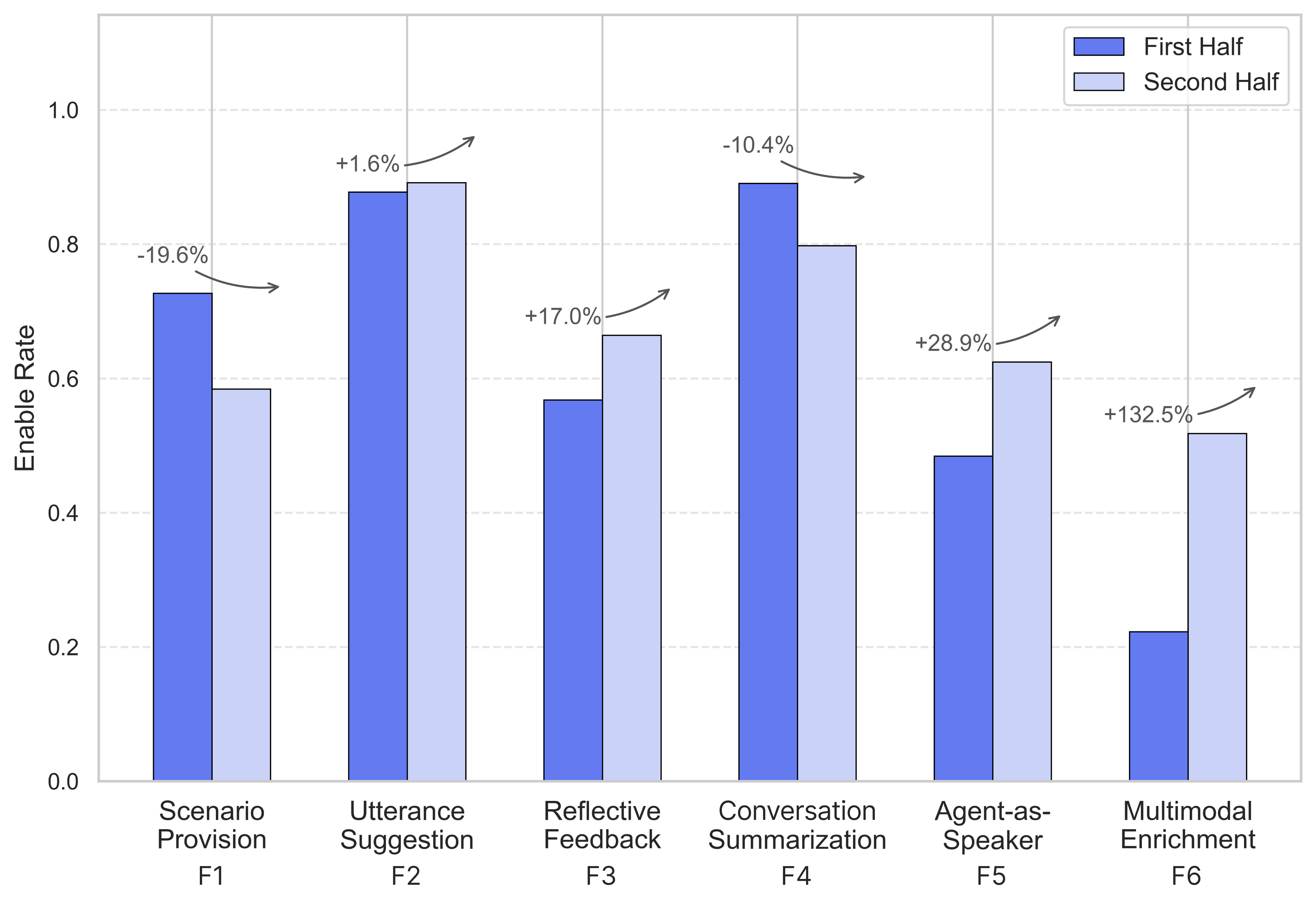} 
        \caption{Function enable rates across the first and second halves of participants' sessions.}  
        \label{fig:Function Enable Rates Half Comparison}  
    \end{minipage}
\end{figure}

\subsubsection{Moving Towards a More Partner-like collaboration mode}
\autoref{figure:Function Enable Rates Bar Chart} shows that \textit{Utterance Suggestion} (F2) and \textit{Conversation Summarization} (F4) were most frequently used, while \textit{Multimodal Enrichment} (F6) was least used; \textit{Scenario Provision} (F1), \textit{Reflective Feedback} (F3), and \textit{Agent-as-Speaker} (F5) showed moderate usage.

Temporal analysis (\autoref{fig:Function Enable Rates Half Comparison}) revealed a significant increase in \textit{Multimodal Enrichment} (F6, $+132.5\%$, $p<0.01$), with smaller gains in \textit{Agent-as-Speaker} (F5, $+28.9\%$) and \textit{Reflective Feedback} (F3, $+17.0\%$), and declines in \textit{Scenario Provision} (F1, $-19.6\%$) and \textit{Conversation Summarization} (F4, $-10.4\%$).

Overall, parents' preferences shifted toward a more partner-like collaboration mode, characterized by reduced reliance on structural support (F1, F4) and increased engagement in co-performance with AI (F5, F6).

\begin{figure}[h]
    \begin{minipage}[t]{0.26\textwidth}
        \centering
        \includegraphics[width=\textwidth]{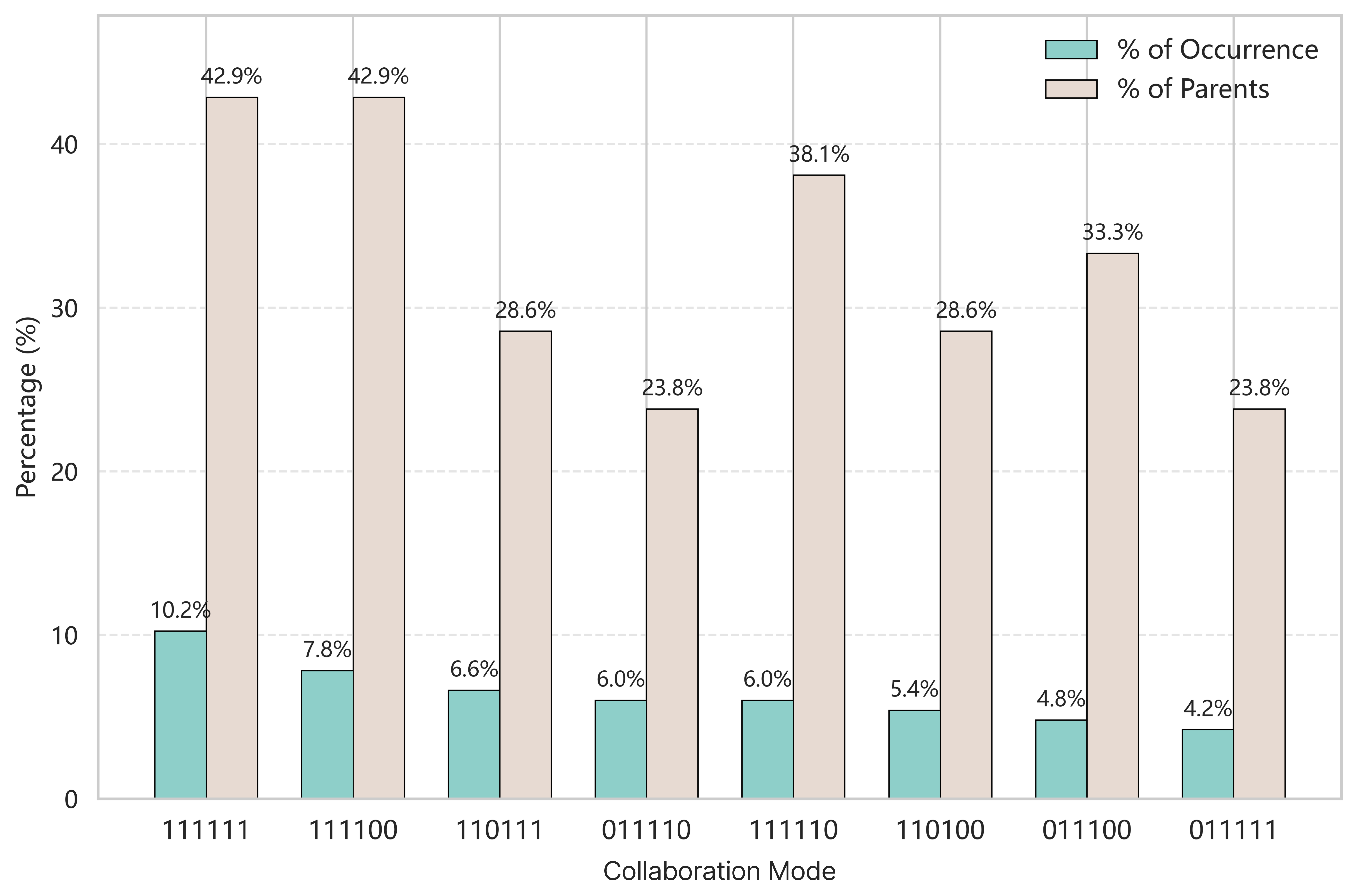}
        \caption{Distribution of the eight most frequent collaboration modes observed in our study.}
        \label{figure:Top 8 Pattern Distribution}
    \end{minipage}
    \hfill
    \begin{minipage}[t]{0.2\textwidth}
        \centering
        \includegraphics[width=\textwidth]{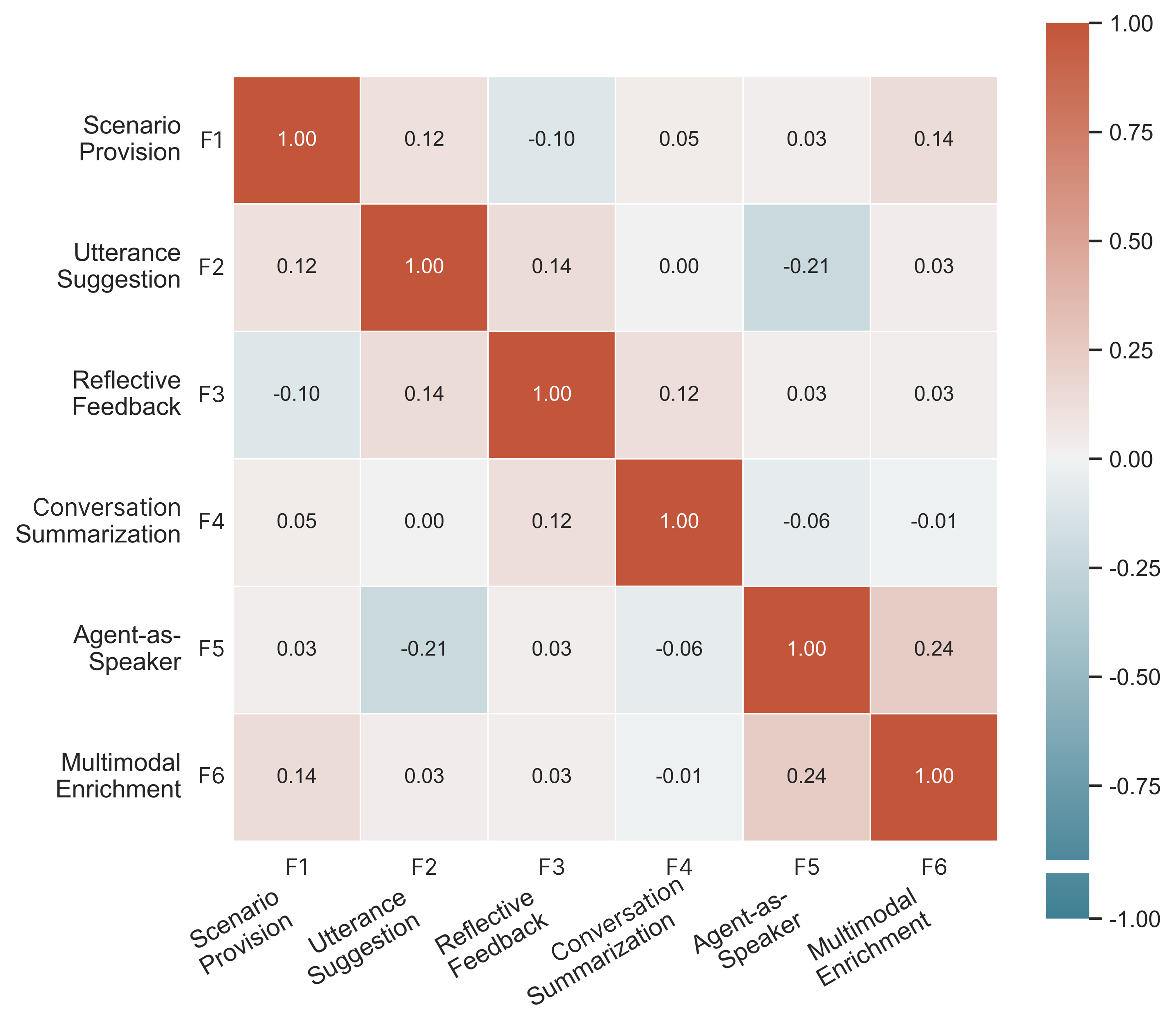} 
        \caption{Heatmap of the correlations between each pair of functions.}  
        \label{fig:Functions Functions Correlation Heatmap}  
    \end{minipage}
\end{figure}

\subsubsection{Adding Reflective or Expressive Support on Top of Basic Assistance}
Parents explored 41 out of 64 possible collaboration modes, indicating flexible needs beyond fixed configurations. Pairwise correlations among functions were weak ($\phi \in [-0.2, 0.2]$; \autoref{fig:Functions Functions Correlation Heatmap}), suggesting that functions were combined largely independently.

Despite this diversity, usage was highly skewed: the top eight modes accounted for 51.20\% of all cases (\autoref{figure:Top 8 Pattern Distribution}). These modes reveal consistent structural patterns:

\textbf{Basic assistance as a constant backbone.} \textit{Utterance Suggestion} (F2) and \textit{Conversation Summarization} (F4) were always enabled (8/8), with \textit{Reflective Feedback} (F3) frequently co-occurring (6/8).

\textbf{Reflection as compensation.} When \textit{Scenario Provision} (F1) was absent, \textit{Reflective Feedback} (F3) was consistently used (3/3).

\textbf{Expressive support as co-performing signal.} \textit{Multimodal Enrichment} (F6) only appeared with \textit{Agent-as-Speaker} (F5) (3/3), indicating that richer expression was adopted when AI took a more active role.

Overall, parents anchored on basic assistance (F2, F4) and selectively layered reflective (F3) or expressive support (F5, F6), revealing a structured yet flexible strategy for configuring collaboration.

\subsection{Contextual Factors Influence Collaboration Modes (RQ2)}
\begin{figure*}[htb] 
\centering
\includegraphics[width=1.0\textwidth]{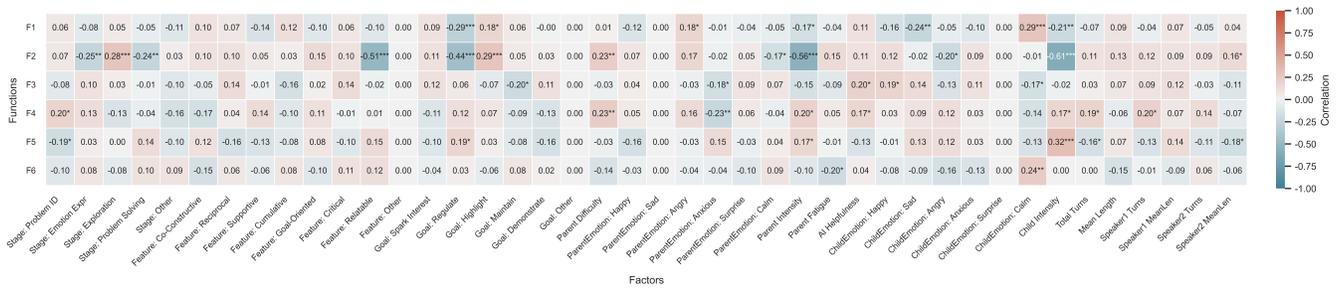}
\caption{Heatmap of correlations between functions and contextual factors, using point-biserial (binary–continuous) and Phi (binary–binary) coefficients. Statistical significance is indicated by asterisks (***: $p<0.001$, **: $p<0.01$, *: $p<0.05$).}
\label{figure:Function-Factor Correlation Heatmap}
\end{figure*}

We examined how contextual factors shape parents' collaboration modes, drawing on Parental Goal Theory \cite{dix1986social, hastings1998parenting}. Specifically, we focus on contextual factors that are observable in lab settings and have been widely discussed in prior work. Factors were grouped into parent-, child-, and relationship-oriented categories (\autoref{tab:contextual factors list}). Correlation analysis (\autoref{figure:Function-Factor Correlation Heatmap}) shows that function usage systematically varies with contextual factors.

\begin{table}[]
\caption{Contextual factors explored in this study. The citations in the ``Data Type'' column refer to prior works that discuss these contextual factors. }
\centering
\resizebox{\columnwidth}{!}{
\begin{tabular}{cccc}
\toprule[1pt]
\textbf{Category}                                                                & \textbf{Contextual Factor} & \textbf{Annotator} & \textbf{Data Type} \\ \hline
\multirow{6}{*}{\begin{tabular}[c]{@{}c@{}}Parent-\\ oriented\end{tabular}}       & Cognitive Load             & Parent             & Numeric \cite{hart1988development}            \\ \cline{2-4} 
                                                                                 & Emotion Type               & Parent             & Single-choice \cite{ekman1992argument}      \\ \cline{2-4} 
                                                                                 & Emotion Intensity          & Parent             & Numeric \cite{roseman1991appraisal}            \\ \cline{2-4} 
                                                                                 & Energy Level               & Parent             & Numeric \cite{hart1988development}            \\ \cline{2-4} 
                                                                                 & Speaker Turns              & AI                 & Numeric            \\ \cline{2-4} 
                                                                                 & Speaker Length             & AI                 & Numeric            \\ \hline
\multirow{4}{*}{\begin{tabular}[c]{@{}c@{}}Child-\\ oriented\end{tabular}}        & Emotion Type               & Parent             & Single-choice\cite{ekman1992argument}      \\ \cline{2-4} 
                                                                                 & Emotion Intensity          & Parent             & Numeric \cite{roseman1991appraisal}            \\ \cline{2-4} 
                                                                                 & Speaker Turns              & AI                 & Numeric            \\ \cline{2-4} 
                                                                                 & Speaker Length             & AI                 & Numeric            \\ \hline
\multirow{5}{*}{\begin{tabular}[c]{@{}c@{}}Relationship-\\ oriented\end{tabular}} & Conversation Stage             & Parent             & Single-choice\cite{mercer2007dialogue, graesser2005autotutor}      \\ \cline{2-4} 
                                                                                 & Conversation Feature           & Parent             & Single-choice \cite{mercer2007dialogue}      \\ \cline{2-4} 
                                                                                 & Conversation Goal              & Parent             & Single-choice \cite{wood1976role}      \\ \cline{2-4} 
                                                                                 & Total Turns                & AI                 & Numeric            \\ \cline{2-4} 
                                                                                 & Conversation Length            & AI                 & Numeric            \\ \bottomrule[1pt]
\end{tabular}
}
\label{tab:contextual factors list}
\end{table}

\subsubsection{Parents' States Lead to Diverse Collaboration Modes}
Parents' internal states strongly shape AI roles.

\textbf{Cognitive load and energy determine support level.} Under high cognitive load, parents prefer direct support, increasing use of \textit{Utterance Suggestion} (F2, $p<0.001$) and \textit{Conversation Summarization} (F4, $p<0.01$). When energy is low, they avoid complex support such as \textit{Multimodal Enrichment} (F6, $p<0.05$).

\textbf{Emotional intensity shifts AI from guide to mediator.} Higher parent emotion intensity reduces acceptance of directive support (F1, F2) but increases reliance on \textit{Conversation Summarization} (F4) and \textit{Agent-as-Speaker} (F5) ($p<0.05$), indicating a shift toward neutral mediation or delegation.

\textbf{Perceived helpfulness enables deeper collaboration.} Higher trust is associated with increased use of reflective and summarizing support (F3, F4; $p<0.05$), reflecting a transition from using AI as a ``tool'' to collaborating with it as a ``partner''.

\subsubsection{Children's Emotional States Lead to Polarized Collaboration Modes}
Children's emotional states induce a polarized pattern.

\textbf{Calm and positive states encourage empowerment.} When children are calm or positive, parents favor \textit{Scenario Provision} (F1, $p<0.001$), \textit{Multimodal Enrichment} (F6, $p<0.01$), and reflection (F3, $p<0.05$). This suggests that in low-stress, open communication environments, parents have more cognitive resources to interact with AI support, such as reference cases or multimedia elements. 

\textbf{High-intensity negative states reduce directive guidance.} When children are sad, angry, or highly aroused, parents decrease use of F1 and F2 ($p<0.01$), avoiding prescriptive suggestions.

\textbf{High emotion intensity shifts AI to front-line support.} With increasing emotion intensity, parents rely more on \textit{Conversation Summarization} (F4) and \textit{Agent-as-Speaker} (F5), using AI as a neutral mediator or substitute speaker to share their burden.

\subsubsection{Relationship-oriented Factors Reshape AI Roles}
Parents' collaboration choices are shaped by relationship-oriented factors, including conversation stage, goal, and interaction intensity.

\textbf{Conversation stage shapes AI support.} Parents rely on AI for sensemaking in the early \textit{Problem Identification} stage (more F4, less F5; $p<0.05$) and for idea generation in the \textit{Exploration} stage (more F2; $p<0.001$), but reduce reliance on content suggestions in the \textit{Emotion Expression} and \textit{Problem Solving} stages (less F2; $p<0.01$).

\textbf{Conversation goals shape whether AI speaks directly.} Under the \textit{Regulate} goal, parents reduce F1/F2 ($p<0.001$) and increase \textit{Agent-as-Speaker} (F5; $p<0.05$), delegating persuasion to AI. Under the \textit{Highlight} goal, they increase F1/F2 ($p<0.05$/$p<0.001$), using AI as a behind-the-scenes supporter to strengthen their own voice.

\textbf{Interaction intensity shapes AI roles.} As conversation turns increase, parents use more \textit{Conversation Summarization} (F4; $p<0.05$) and less \textit{Agent-as-Speaker} (F5; $p<0.05$), positioning AI as an organizer of the conversation rather than a third speaker.

\subsection{Three Types of Strategies Parents Employ to Engage with AI (RQ3)}
Based on interview data, we identify three strategy types guiding parent-AI engagement: \textit{parent-oriented}, \textit{child-oriented}, and \textit{relationship-oriented}. These extend Parental Goal Theory \cite{dix1986social, hastings1998parenting} to AI-supported parent-child interaction.

\subsubsection{Parent-Oriented Strategies}
Parents adopted self-focused strategies to (1) compensate for their limitations and (2) regulate their own cognitive and emotional states.

\textbf{(1) Compensating for limitations and maintaining conversational control.}
11 of 21 parents used AI for real-time feedback and information retrieval to compensate for communication limitations, with eight reporting that AI helped identify deficiencies and support self-correction. As P11 noted, ``\textit{it can point out some parts of my speech that are not quite right... I can adjust my direction accordingly.}'' Parents also relied on AI to address knowledge gaps (e.g., explaining ``new moon'' to children) and to provide psychologically appropriate responses. In emotionally sensitive situations, AI's phrasing suggestions were used to avoid potentially harmful expressions.

9 parents further leveraged AI to improve interaction quality. Seven described using emotional insights to ``shift from evaluative to empathetic communication'' (P18), while six relied on AI's suggestions to maintain conversational flow when unsure how to initiate a topic (P3). 

Under high stress or uncertainty, some parents adopted an ``all-round reliance'' strategy by enabling multiple functions simultaneously to maintain conversational control. As P4 noted, ``\textit{I was worried that after my child expressed something, I would not know how to respond. So I kept Functions 2, 3, and 4 enabled to make sure that after my child spoke, I could immediately pick up her words.}''

\textbf{(2) Managing parents' emotional and cognitive states.}
Parents reported using AI to regulate their internal states during conversations. Three described using reflective prompts to manage agitation and prevent negative emotions (P14: ``\textit{parents' emotions can affect the conversation, and AI can help me with that}''). When distracted by work or other demands, two parents used summarization and prompting to refocus and maintain continuity, noting that the system could ``\textit{help pull me back to the core of the conversation}'' (P7). Additionally, parents used AI to broaden thinking when constrained by habitual patterns, seeking alternative perspectives and inspiration (P2: ``\textit{When I feel like my words are confined to my habitual ways of thinking, I try (using AI) to get some help}'').

\subsubsection{Child-Oriented Strategies}
Parents adapted AI use by continuously observing and supporting their children's behaviors, focusing on (1) monitoring children's status and (2) addressing children's expressive difficulties.

\textbf{(1) Observing and evaluating children's short- and long-term status.}
Parents dynamically adjusted system features based on children's attention. For instance, some enabled or avoided multimodal features depending on whether they supported or distracted engagement, noting that ``\textit{the child might start talking to the AI, which would distract his attention}'' but could also ``\textit{raise his interest}'' (P4). 

Parents also relied on system feedback and scenario-based prompts to better perceive children's emotions. They report that system feedback can help them perceive their child's feelings more sensitively. Some parents further highlight that \textit{Scenario Provision)} (F1) allows them to observe the child's emotions and understand their ``\textit{first reaction}'' and ``\textit{true thoughts}'' (P4).

Beyond moment-to-moment interaction, parents considered long-term factors such as age, interests, and personality. They adjusted feature use accordingly---for example, reducing ``childish'' cases for older children (P9) or leveraging familiar scenarios to help children ``\textit{immerse more easily}'' (P3) and ``\textit{better control their emotions}'' (P19). These practices reflect a child-adaptive strategy of tailoring AI support to developmental and individual needs.

\textbf{(2) Coping with children's expression barriers.}
When children showed resistance or struggled to express themselves, parents used AI as a ``\textit{backup guide}'' to sustain interaction. They enabled \textit{Utterance Suggestion} to restart stalled conversations (P3) or used multimodal cues to redirect attention (P1). Parents also leveraged AI to regulate conversational pacing. Several parents highlighted that \textit{Reflective Feedback} (F3) is particularly effective in slowing down the interaction. Pauses for self-reflection ``\textit{give the child more space to organize their language}'' (P7).

To stimulate expression, parents employed AI-generated cases, prompts, or scenarios to elicit responses. For example, reading AI-suggested cases helped children ``\textit{align their thinking}'' (P4), while prompts encouraged children ``\textit{be willing to carry the conversation more deeply}'' (P6). In some cases, multimodal content was used at the end of conversations to foster emotional resonance and extend thinking (P4).

\subsubsection{Relationship-Oriented Strategies}
Beyond parent- or child-focused strategies, parents positioned COMPASS as a relational mediator that both directly intervenes in conversation and indirectly shapes interaction.
 
\textbf{(1) Direct AI involvement as a mediator.}
Parents used AI as a neutral third party to prompt children's expression and reinforce their own perspectives. For example, AI could ``\textit{guide the child to express inner thoughts, just like a psychological expert}'' (P19). AI could also repeat parental viewpoints (P17:``\textit{it made my opinion more convincing to the child }'').

Some also relied on AI to sustain interaction, keeping it active to facilitate turn-taking and help children elaborate incomplete responses (P4:``\textit{ Adding Function 5 helped inspire my child to say more and finish their sentences }''). 

AI further functioned as an emotional buffer in tense situations, helping both sides regulate emotions and continue communication. As one parent put it: ``\textit{The biggest function (of COMPASS) is to control emotions.}'' (P19). 

\textbf{(2) Indirect influence on relationship dynamics.}
Parents described COMPASS as shaping interaction climate through subtle, indirect influence. Its presence encouraged more serious and reflective engagement (P17: ``\textit{it can help both my child and me take (this topic) more seriously}''). 

AI-generated cues also fostered moments of closeness and responsiveness. Parents reported that AI-generated prompts helped foster intimacy in ways they might not have achieved on their own. For example, after COMPASS generated an illustration for a storybook, one parent recalled: ``\textit{My child sat on my lap, just like in the AI storybook, and we had a close physical contact. }'' (P1). 

Finally, parents used AI to check whether they and their children were ``on the same page.'' Real-time summaries (F4) reassured them about alignment: ``\textit{I want to check our communication is (happening)... (whether) we are engaging in effective communication}'' (P20).
\section{Discussion}

In this work, we use COMPASS as a research probe to investigate how parents configure AI involvement in real-time conversations with children. Our findings reveal that parent–AI collaboration unfolds as a dynamic, goal-driven, and exploratory process over time. Here, we further discuss (a) how collaboration modes should be conceptualized as adaptive computational entities, (b) how parental goal theory is reshaped in AI-engaged parent–child conversations, and (c) how these insights translate into design implications for parent–AI collaborative systems. Together, these findings inform the design of future family-based systems that enable computable, adaptive collaboration in open-ended educational interactions.

\subsection{Toward Computable Adaptive Collaboration in Parent-AI Systems}
Prior family-related AI systems have highlighted the importance of flexible AI participation, such as through adjustable parental involvement level or shared role-taking \cite{zhang2022storybuddy, vargas2023talemate}. However, these systems largely operationalize collaboration as a set of predefined roles and participation options.

Our findings extend this line of work by showing that, in sensitive parent-child conversations, collaboration cannot be characterized as a fixed allocation of roles and responsibilities. Instead, parent-AI collaboration modes are continuously recomputed through ongoing trade-offs among multiple parental goals. This dynamic reconfiguration was substantial in our study: parents switched modes an average of 7.90 times per conversation and explored 41 of the 64 possible function combinations. By repeatedly recombining AI functions, parents continuously reconfigured AI's role, form, and level of involvement in response to evolving conversational contexts.

Besides, this recomputation unfolds through a progressive and exploratory process. Parents typically began with backstage support to maintain conversational control, and gradually granted AI more visible, partner-like roles. This progression was driven by parents’ need to assess AI controllability and safety before incorporating it into the sensitive parent–child interaction space, reflecting a trust-building process. At the same time, not all collaboration modes were equally sustained: while some modes stabilized into recurring patterns, others appeared as short-lived transitional states. These transitional modes reflect parents’ ongoing experimentation and familiarization with AI capabilities, suggesting that collaboration emerges through iterative exploration rather than immediate convergence to fixed strategies.

\textbf{These findings suggest that, in parent-AI systems, collaboration modes should be explicitly modeled as computational entities that can be continuously inferred and updated}. Specifically, collaboration modes can be computed from observable contextual signals (e.g., conversational stage, child state, parental state) to support real-time balancing among competing parental goals, while also adapting to parents’ evolving trust and familiarity with AI. This extends prior HCI work on context-aware and adaptive human-AI interaction systems \cite{amershi2019guidelines, mcgrath2025collaborative}. 

\textbf{Building on this insight, we propose a new design direction: family-related interfaces that enable computable adaptive collaboration}. Rather than treating collaboration modes as static interface settings, such systems could continuously sense interactional context and adaptively determine how, when, and to what extent AI participates, enabling fluid transitions across collaboration modes. This perspective may also inform the design of human-AI collaboration systems in other sensitive and open-ended HCI domains such as education \cite{mei2025pacee, chen2025cograder, li2025designing}, healthcare \cite{jo2025decicare, xu2022towards}, and counseling \cite{wang2025practice, shah2022modeling, hsu2025helping}.

\subsection{Revisiting Parental Goal Theory in AI-Engaged Parent-Child Conversation}
Parenting goals are typically classified as parent-oriented, child-oriented, and relationship-oriented goals \cite{dix1986social, hastings1998parenting}. Grounded in our findings, we revisit these categories and show how they are reshaped in AI-engaged parent-child conversation contexts.

\subsubsection{Reconceptualizing Parent-Oriented Goals to Include AI-Mediated Orchestration}
Prior work shows that parents aim to maintain competence, bridge knowledge gaps, and regulate their own cognitive and emotional states when interacting with children \cite{coleman1998self, deater1998parenting, heerman2017parenting}. While our findings support these parent-oriented goals, we extend this view by introducing \textit{AI-mediated orchestration}: in AI-engaged parent–child conversations, parent-oriented goals are no longer confined to self-regulation, but involve actively configuring AI participation and continuously aligning parents' mental models of their children.

First, parent-oriented goals expand from self-regulation to the regulation of AI use. Parents act as adaptive orchestrators who continuously determine how, when, and to what extent AI participates in interaction. Rather than using AI as a fixed tool, they iteratively explore system capabilities and dynamically reconfigure collaboration modes, often grounded in their daily technology-use habits. This reframes parent-oriented goals to include managing the AI-mediated interaction process. Moreover, how parents regulate AI use is shaped by their AI literacy (e.g., familiarity and proficiency), which in turn influences how they pursue different levels of AI involvement.

Second, parent-oriented goals extend to the AI-mediated construction of child understanding. Traditionally, parents rely on partial observations filtered through prior beliefs to interpret children's emotions and behaviors. AI introduces in-situ interpretations of children's moment-to-moment states (e.g., emotion type, intensity, and expression patterns), which parents appropriate to reflect on and recalibrate their own judgments. In our study, parents actively integrate these AI-generated interpretations into their mental models of their children, which indirectly supports child-oriented and relationship-oriented goals during interaction.

\subsubsection{Child-Oriented Goals as Long-Term Development and In-Situ Regulation}
While our system does not include AI functions for direct child-orientated support, it supports parents to more effectively pursue their own child-oriented goals within the conversation. Therefore, it remains necessary to examine how AI support reshapes parents’ conceptualizations of child-oriented goals.

Child-oriented goals have traditionally focused on children's long-term cognitive and socio-emotional development \cite{chang2017parenting, dix2014parenting, ho2022measuring, sun2024exploring}, with AI systems primarily supporting learning, emotion recognition, and self-expression \cite{sun2024exploring, chen2024storysparkqa, chen2025characterizing}. Our findings extend this view by showing that child-oriented goals also involve attending to children's moment-to-moment states (e.g., emotional type, intensity, and expression), a dimension that remains underexplored in prior work. While parents are known to regulate children's emotions in interaction \cite{eisenberg2004emotion, eisenberg2002children}, AI reshapes this process through real-time monitoring of children's states. 
This reframes child-oriented goals as the integration of long-term developmental objectives and in-situ regulation, highlighting a new design space for AI systems to support and dynamically balance these two aspects.
 
\subsubsection{Reconfiguring Relationship-Oriented Goals as Triadic Relationship Orchestration}
Prior work frames relationship-oriented goals as maintaining and enhancing the quality of the parent-child relationship \cite{laursen2009parent, laursen2003parent}. Our findings support this view; however, we further reveal that in AI-engaged parent-child conversations, relational goals are no longer confined to dyadic maintenance. Instead, they are reconfigured as the orchestration of a triadic relationship among parent, child, and AI.

AI did not merely provide instrumental assistance during conversation. Parents positioned it as a relational mediator that could sustain interaction, encourage children's disclosure, and buffer emotional tension---particularly under high emotional load, when parents struggled to regulate conversations without escalating conflict \cite{barros2015parental, morris2007role}. In such moments, AI helped preserve conversational continuity and created a less confrontational interactional space, allowing parents to maintain connection with the child while navigating difficult emotional exchanges.

These findings reconfigure relationship-oriented goals from improving the parent-child bond alone to managing how AI participates in the relationship: when it should act as a neutral mediator, when it should remain in the background, and the degree of relational influence it should be granted. As a result, relationship-oriented goals expand to encompass the coordination of participation, authority, and intimacy across the parent-child-AI triad.

Moreover, parents' trust in AI and their attitudes toward its social role shape the degree of relational authority they delegate to it. This introduces an ongoing negotiation among AI participation, parental authority, and children's autonomy. Maintaining a strong parent-child relationship thus increasingly depends on how parents manage AI's relational role, rather than simply on AI’s presence.

\subsection{Design Implications}
Grounded in our findings, we derive design implications for designing parent–AI systems that support flexible collaboration in parent-child interactions.

\subsubsection{Context-aware Explainability without Disrupting Interaction}
AI interventions in sensitive contexts can appear intrusive without explanation, undermining parental trust \cite{binns2018s, kulesza2015principles}. Our results show that parents' collaboration preferences shift systematically with emotional intensity, conversation stage, and parental goals. Systems should therefore provide context-aware explanations that clarify strategy–context links \cite{wang2019designing, dwivedi2023explainable} in family communication settings. 

However, continuous explanations risk cognitive overload and interrupt the smooth flow of parent-child conversation. Explainability should be treated not as a constant stream of information, but as an adaptive interactional feature, where timing, modality, and granularity are dynamically adjusted to balance transparency with conversational continuity.

\subsubsection{Distinguishing Exploration from Stable Usage Patterns}
Parents engaged with the system through an iterative process of trial and adjustment, often experimenting with new functions or revisiting previously dismissed ones. This means that system logs do not always reflect stable preferences. Therefore, when modeling user preferences \cite{pi2020search, jawaheer2014modeling}, systems should distinguish exploration from stable usage patterns. 

Moreover, rather than treating exploration as noise, it should be recognized as a valuable meta-strategy that reflects how parents probe AI capabilities. Future systems may leverage such patterns to personalize support based on users' exploration styles.

\subsubsection{Rethinking AI Literacy in Parenting}
Collaboration flexibility varied with parents' AI literacy: high-literacy parents dynamically reconfigured functions, whereas low-literacy parents relied on fixed collaboration modes, suggesting that current systems may impose excessive complexity and cognitive burden on low-literacy parents. This highlights the need for literacy-aware designs that adapt system complexity to parents' capabilities. In parenting contexts, however, AI literacy extends beyond technical proficiency \cite{long2020ai, ng2021conceptualizing} to include relational sensitivity---knowing when and how to involve AI without disrupting the parent–child relationship. At the same time, overly simplified designs risk constraining parents' agency and hindering long-term learning. Building on prior work on fostering AI literacy \cite{lee2021developing, ng2024fostering}, these findings point to balancing immediate usability with the gradual development of parents' AI literacy.

\subsubsection{Context-Aware Restraint in Sensitive Parenting Moments}
Parents adapted AI use to context: in calm, cognitively oriented moments, they preferred structured support, whereas in emotionally intense situations, they avoided directive supports and relied on alignment and mediation. While prior HCI work highlights context-aware support \cite{ogata2004context, chen2023gap, chen2025investigating}, we further identify the need for context-aware restraint. AI should not only know when to step in but also hold back in ``no-go'' moments (e.g., avoiding rigid responses when children are sad). This aligns with ``calm technology'' \cite{weiser1996designing, weiser1997coming} and reframes deliberate AI silence as a form of positive intelligence that protects sensitive parent–child interactions.

\subsection{Limitations and Future Work}
First, participants were limited to Chinese parents and children, with most parents being mothers, reflecting common caregiving patterns \cite{mcbride1993comparison, schoppe2013comparisons, hernawati2020differences}. Future work should examine broader cultural contexts and parental roles, including mother–father differences in AI collaboration under similar structured interactions.

Second, the lab-based setting ensured controlled observation of moment-to-moment collaboration shifts but may not reflect everyday home communication routines. Longitudinal in-home deployments are needed to understand how parent–AI collaboration evolves in natural family settings.

Third, we did not assess children's outcomes, as COMPASS was designed as a probe to surface parental strategies rather than a full intervention. Our contribution lies in theorizing and understanding parent-AI collaboration. Future research should further investigate how different parent–AI divisions of labor influence children's communication and development.

Finally, COMPASS focused on six common AI support functions identified in our formative study, which do not cover the full design space. While this scope is sufficient for our purpose of examining parents' shifting collaboration needs, future work could extend parent–AI collaboration to provide richer, more personalized support.

\section{Conclusion}
We developed COMPASS, a research probe designed to provide flexible AI support for parents during parent-child conversations.
Using COMPASS as a probe in a study with 21 parent-child pairs, we find that parent–AI collaboration evolves from backstage support to partner-style co-performance, shifting with contextual factors to balance cognitive and emotional regulation.
We identify three parental strategies, namely parent-oriented, child-oriented, and relationship-oriented strategies, extending Parental Goal Theory and informing the design of future parental support systems that enable flexible and effective partnerships.

\bibliographystyle{ACM-Reference-Format}
\bibliography{ref/main_citation}

\clearpage
\appendix
\newpage

\section{Details of the Six Parent Support Functions}
\label{appendix:Details of the Six Parent Support Functions}
This section presents the six AI support functions (F1-F6) identified from parents’ requests during the formative study.

\textbf{Scenario Provision (F1).} 
Parents preferred that the AI introduce topics through contextualized storytelling or illustrative examples to evoke children's empathy and understanding, rather than directly providing fixed utterances. The ability of the AI to generate stories naturally aligns with everyday communication and children's cognitive development stages, making it ``\textit{easier for children to understand and accept}'' (P6).

\textbf{Utterance Suggestion (F2).} 
Seven out of nine parents expected the AI to provide concise, contextually appropriate utterance suggestions that they ``\textit{could read aloud directly}'' (P8) at critical moments to help them overcome difficulties in phrasing. This highlights parents' need for immediacy and contextual adaptability rather than reliance on generic templates.

\textbf{Reflective Feedback (F3).} 
Some parents emphasized their preference for heuristic rather than judgmental feedback: ``\textit{I don't want the AI to criticize me in a harsh tone... I need it to inspire me on how to think}'' (P2). Parents were more inclined to view the AI as a gentle thinking partner rather than an authoritative instructor, and therefore, the reflection function should focus on guidance and empathy.

\textbf{Conversation Summarization (F4).} 
Six parents reported that they often found it difficult to track their children's emotions and perspectives, and thus hoped that the AI could summarize both parties' positions and feelings in real time. One parent expressed a desire for the AI to analyze family members' thoughts and reasoning, ``\textit{so that everyone can understand each other, and I can respond to my child better}'' (P1). This reflects parents' expectations for the AI to function as an information filter and emotional mediator, helping maintain smooth conversation and promote understanding.

\textbf{Agent-as-Speaker (F5).} 
Although most parents preferred the AI to act as an assistant, some welcomed its brief involvement as a ``\textit{third-party friend}'' or ``\textit{story character}'' when conversations stalled (P4). The premise was that ``\textit{the AI should be a good listener}'', intervening naturally and gently without interrupting parent–child interaction (P2).

\textbf{Multimodal Enrichment (F6).} 
Some parents noted that children aged 6-8 have difficulty understanding abstract vocabulary, and therefore suggested that the AI provide multimodal feedback such as picture book illustrations, enabling children to ``\textit{be more willing to continue the conversation}'' (P5). This reflects parents' desire for the system to reduce comprehension burden through concrete and vivid forms while maintaining engagement in the conversation.

\section{Implementation of the COMPASS Research Probe}
\label{appendix:Implementation of the COMPASS Research Probe}
We implemented the main system workflow on a Flask server, while the frontend interface is built in Vue and communicates with the server via WebSocket. The frontend runs on a touchscreen-enabled laptop to provide an intuitive and interactive interface for parents. For the \textit{Audio Processor}, we utilized the TINGWU API, with audio sliced and compressed on the frontend prior to transcription. We also employed GPT-4o-mini to map each diarized speaker index to the corresponding role (parent or child). For all six functions in the Function Library, we leveraged the GPT-4o API to balance speed and output quality. For the \textit{Multimodal Enrichment Module}, we employed the Stable Diffusion XL (SDXL) model due to speed considerations, whereas for the \textit{Agent-as-Speaker Module}, we used Alibaba Cloud's streaming synthesis API configured with a gentle female voice profile.

\section{Additional User Evaluation Details}
\label{appendix:Additional User Evaluation Details}
This section provides participant demographic details for the user evaluation (\autoref{tab:user-study-demographic}).

\begin{table}[htb]
\caption{Participant demographic information in the user study.}
\centering
\small
\begin{tabular}{cccccc}
\toprule[1pt]
\textbf{ID} & \textbf{\begin{tabular}[c]{@{}c@{}}Child\\ Age\end{tabular}} & \textbf{\begin{tabular}[c]{@{}c@{}}Child\\ Gender\end{tabular}} & \textbf{\begin{tabular}[c]{@{}c@{}}Parent\\ Age\end{tabular}} & \textbf{\begin{tabular}[c]{@{}c@{}}Parent\\ Gender\end{tabular}} & \textbf{\begin{tabular}[c]{@{}c@{}}Parental\\ Education\end{tabular}} \\ \hline
1           & 6                                                            & Male                                                            & 36                                                            & Female                                                           & Bachelor's                                                            \\ \hline
2           & 7                                                            & Male                                                            & 38                                                            & Female                                                           & Master's                                                              \\ \hline
3           & 7                                                            & Male                                                            & 43                                                            & Female                                                           & Bachelor's                                                            \\ \hline
4           & 8                                                            & Female                                                          & 37                                                            & Female                                                           & Bachelor's                                                            \\ \hline
5           & 6                                                            & Female                                                          & 42                                                            & Male                                                             & Ph.D.                                                                 \\ \hline
6           & 7                                                            & Female                                                          & 32                                                            & Female                                                           & Ph.D.                                                                 \\ \hline
7           & 6                                                            & Female                                                          & 33                                                            & Female                                                           & Master's                                                              \\ \hline
8           & 7                                                            & Male                                                            & 45                                                            & Female                                                           & Master's                                                              \\ \hline
9           & 8                                                            & Female                                                          & 38                                                            & Female                                                           & Master's                                                              \\ \hline
10          & 7                                                            & Male                                                            & 41                                                            & Male                                                             & Master's                                                              \\ \hline
11          & 7                                                            & Female                                                          & 38                                                            & Female                                                           & Bachelor's                                                            \\ \hline
12          & 8                                                            & Female                                                          & 42                                                            & Female                                                           & Bachelor's                                                            \\ \hline
13          & 6                                                            & Male                                                            & 47                                                            & Male                                                             & Master's                                                              \\ \hline
14          & 7                                                            & Male                                                            & 33                                                            & Female                                                           & Master's                                                              \\ \hline
15          & 8                                                            & Male                                                            & 42                                                            & Female                                                           & Bachelor's                                                            \\ \hline
16          & 7                                                            & Female                                                          & 35                                                            & Female                                                           & Master's                                                              \\ \hline
17          & 6                                                            & Female                                                          & 40                                                            & Female                                                           & Master's                                                              \\ \hline
18          & 7                                                            & Male                                                            & 42                                                            & Male                                                             & Master's                                                              \\ \hline
19          & 6                                                            & Female                                                          & 36                                                            & Female                                                           & Bachelor's                                                            \\ \hline
20          & 8                                                            & Male                                                            & 37                                                            & Female                                                           & Bachelor's                                                            \\ \hline
21          & 8                                                            & Male                                                            & 43                                                            & Female                                                           & Associate's                                                           \\ \bottomrule[1pt]
\end{tabular}
\label{tab:user-study-demographic}
\end{table}

% \section{Co-occurrence of the Scaffolding Functions} 
% \label{appendix:Co-occurrence of the Scaffolding Functions}
% Overall, the phi coefficients among the six functions largely fell within the range of $-0.2$ to $0.2$, indicating that parents flexibly and independently combined functions according to different contexts, without strong tendencies toward fixed pairings (see Figure~\ref{fig:Functions Functions Correlation Heatmap}).

% The most notable positive association emerged between \textit{Agent-as-Speaker} (F5) and \textit{Multimodal Enrichment} (F6) ($\phi=0.24$). This suggests that when parents were willing to let the AI directly participate in conversations with their child, they were also more inclined to incorporate multimodal image-text support.

% In contrast, the strongest negative association was observed between \textit{Utterance Suggestion} (F2) and \textit{Agent-as-Speaker} (F5) ($\phi=-0.21$). This points to a potential trade-off: some parents preferred to remain in control of the conversation with AI providing background prompts, while others delegated more agency to the AI by allowing it to directly engage with the child. 

\end{document}